\documentclass[11pt,a4paper]{article}

\usepackage[utf8]{inputenc}
\def\papertitle{\bf Final-state rescattering in  $\bar{B}^{0}_{(s)}\to \Lambda^{+}_{c}\bar{\Lambda}^{-}_{c}$ decays} 
\usepackage{authblk}

\title{\papertitle}
\author[1,2]{Zhu-Ding Duan,\footnote{Email: duanzd@mail.imu.edu.cn}}
\author[1]{\,Xiao Huang,\footnote{Email: huangxiao2025@lzu.edu.cn, corresponding author}}
\author[1]{\,Dong-Hao Li,\footnote{Email: lidh@lzu.edu.cn, corresponding author}}
\author[2]{\,Run-Hui Li,\footnote{Email: lirh@imu.edu.cn, corresponding author}}
\author[1,4]{\,Jian-Peng Wang,\footnote{Email: wangjp20@lzu.edu.cn, corresponding author}}
\author[1,3]{and Fu-Sheng Yu\footnote{Email: yufsh@lzu.edu.cn, corresponding author}}

\affil[1]{Frontiers Science Center for Rare Isotopes, and School of Nuclear Science and Technology, Lanzhou University, Lanzhou 730000, China}
\affil[2]{School of Physical Science and Technology, Inner Mongolia Key Laboratory of Microscale Physics and Atomic Manufacturing, Inner Mongolia University, Hohhot 010021, China}
\affil[3]{Lanzhou Center for Theoretical Physics, Key Laboratory of Theoretical Physics of Gansu Province, Key Laboratory of Quantum Theory and Applications of MoE, Gansu Provincial Research Center for Basic Disciplines of Quantum Physics, Lanzhou University, Lanzhou 730000, China}
\affil[4]{Physik Department, Universität Siegen, Walter-Flex-Str.3, D-57068 Siegen, Germany}
\usepackage{changes}
\usepackage[top=1in, bottom=1.25in, left=1in, right=1in]{geometry}
\usepackage{microtype}
\usepackage{lineno}
\usepackage{xspace}
\usepackage{booktabs}
\usepackage{comment}
\usepackage{slashed}

\usepackage{caption}

\usepackage{graphicx}  

\usepackage{color}
\usepackage{colortbl}
\DeclareGraphicsExtensions{.pdf,.PDF,png,.PNG}

\usepackage{amsmath} 
\usepackage{amsfonts,amssymb,amsthm,bm,braket,yhmath}
\usepackage{upgreek} 

\usepackage{hyperref}
\usepackage{hyperxmp}
\usepackage[figuresright]{rotating}
\usepackage{hypcap} 
\hypersetup{hidelinks} 
\usepackage{txfonts}
\usepackage{amssymb}
\usepackage{cases}
\usepackage{algorithm}
\usepackage{algpseudocode}
\usepackage{mathrsfs}
\usepackage{color}
\usepackage{amscd,amssymb,amsthm,amsmath,bm,graphicx,psfrag}
\usepackage{epsfig,mathrsfs}
\usepackage[bf,SL,BF]{subfigure}
\usepackage{amsmath}
\usepackage{pict2e,keyval,fp}
\usepackage{booktabs}
\usepackage{array}
\usepackage{multirow}

\theoremstyle{definition}

\numberwithin{equation}{section}

\def\kaon   {{\ensuremath{K}}\xspace}
\def\Kbar    {{\kern 0.2em\overline{\kern -0.2em \kaon}{}}\xspace}

\def\KorKbar    {\kern 0.18em\optbar{\kern -0.18em K}{}\xspace}

\def\Dbar    {{\kern 0.2em\overline{\kern -0.2em D}{}}\xspace}

\def\DorDbar    {\kern 0.18em\optbar{\kern -0.18em D}{}\xspace}

\def\Lbar        {{\ensuremath{\kern 0.1em\overline{\kern -0.1em\Lambda}}}\xspace}
\def\LorLbar    {\kern 0.18em\optbar{\kern -0.18em \PLambda}{}\xspace}

\newcommand{\tev}{\ensuremath{\mathrm{\,Te\kern -0.1em V}}\xspace}
\newcommand{\gev}{\ensuremath{\mathrm{\,Ge\kern -0.1em V}}\xspace}
\newcommand{\mev}{\ensuremath{\mathrm{\,Me\kern -0.1em V}}\xspace}
\newcommand{\kev}{\ensuremath{\mathrm{\,ke\kern -0.1em V}}\xspace}
\newcommand{\ev}{\ensuremath{\mathrm{\,e\kern -0.1em V}}\xspace}
\newcommand{\gevc}{\ensuremath{{\mathrm{\,Ge\kern -0.1em V\!/}c}}\xspace}
\newcommand{\mevc}{\ensuremath{{\mathrm{\,Me\kern -0.1em V\!/}c}}\xspace}
\newcommand{\gevcc}{\ensuremath{{\mathrm{\,Ge\kern -0.1em V\!/}c^2}}\xspace}
\newcommand{\gevgevcccc}{\ensuremath{{\mathrm{\,Ge\kern -0.1em V^2\!/}c^4}}\xspace}
\newcommand{\mevcc}{\ensuremath{{\mathrm{\,Me\kern -0.1em V\!/}c^2}}\xspace}

\usepackage{cite} 
\usepackage{mciteplus}

\begin{document}
\maketitle
\begin{abstract}
The LHCb Collaboration has recently reported the first observation of the decay \(\bar B_s^0\to \Lambda_c^+\bar\Lambda_c^-\), along with measurements of the branching fractions for both \(\bar B^0\to \Lambda_c^+\bar\Lambda_c^-\) and \(\bar B_s^0\to \Lambda_c^+\bar\Lambda_c^-\). In this work, we investigate these two decays within the framework of final state re-scattering. Our results show that the predicted branching fractions are consistent with the experimental measurements, which indicates the significant role of long-distance final-state interactions in such baryonic $B$ decays. Furthermore, we present predictions for the direct CP asymmetries and the asymmetry parameters. Numerically, both decays exhibit nearly vanishing direct CP asymmetries and suppressed longitudinal polarizations $\alpha$, whereas the sizable values of $\beta$ and $\gamma$ reflect strong interference between the two helicity amplitudes and provide sensitive probes of the decay dynamics.

\newpage
\end{abstract}
\section{Introduction}
\begin{figure}[H]
\centering 
     \includegraphics[width=12cm,height = 3.7cm]
     {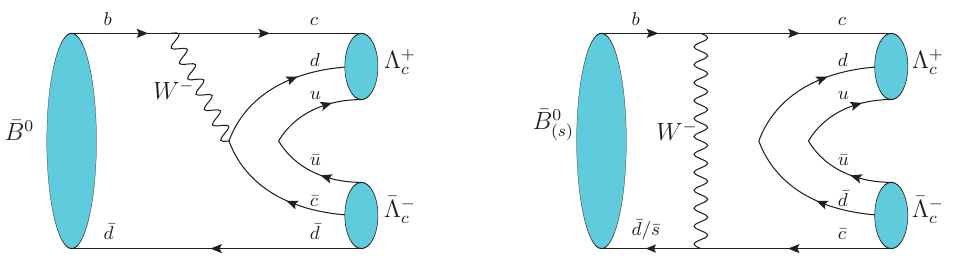}
\caption{The internal W-emission diagram (left) and W-exchange diagram (right).}\label{Topological diagram}
\end{figure}

The recent observation by the LHCb Collaboration of $\bar B_s^0 \to \Lambda_c^+\bar{\Lambda}_c^-$, together with evidence for 
$\bar B^0 \to \Lambda_c^+\bar{\Lambda}_c^-$, furtherly inspires theoretical studies. The branching fractions have been measured as 
\begin{align}
\mathcal{B}(\bar B_s^0\to \Lambda_c^+\bar{\Lambda}_c^-)
&=(5.0\pm1.3\pm0.5\pm0.8)\times10^{-5},\\
\mathcal{B}(\bar B^0\to \Lambda_c^+\bar{\Lambda}_c^-)&=(1.01^{+0.27}_{-0.28}\pm0.08\pm0.15)\times10^{-5},
\end{align}
respectively~\cite{LHCb:2025ueu}. Based on the quark diagram analysis, $\bar B^0 \to \Lambda_c^+\bar{\Lambda}_c^-$ receives contributions from both internal W-emission and W-exchange topologies, whereas $\bar B_s^0 \to \Lambda_c^+\bar{\Lambda}_c^-$ proceeds only through W-exchange topology, as shown in Fig.~\ref{Topological diagram}. The W-exchange contributions have usually been regarded as helicity suppressed and were therefore often regarded as negligible in phenomenological analyses of two-body baryonic $B$ meson decays~\cite{Cheng:2002sa,Cheng:2009yz}. Within naive diagrammatic estimation, the decay $\bar B^0 \to \Lambda_c^+\bar{\Lambda}_c^-$ is generally expected to be dominated by W-emission amplitude, while $\bar B_s^0 \to \Lambda_c^+\bar{\Lambda}_c^-$ is largely suppressed. Experimentally, the branching fractions of two decays however are at the same order. This indicates that the W-exchange contribution is sizable and cannot simply be treated as a suppressed and ignored effect.

For charmful baryonic $B$ decays, the energy release is small, and hence the helicity suppression can be significantly alleviated by heavy-quark mass effects. As a result, the W-exchange contribution may be enhanced through long
distance final-state soft re-scattering effects, which are not included in the naive factorization estimation. An analysis based on $\text{SU}(3)$ flavor symmetry pointed out that the naive single-W-emission picture tends to overestimate the branching fraction of $\bar B^0 \to \Lambda_c^+\bar{\Lambda}_c^-$, and that a non-negligible W-exchange contribution is necessary for alleviating the tension between theory and experiment~ \cite{Hsiao:2023mud}.  Ref.~\cite{Hsiao:2023mud} predicts
$\mathcal{B}(\bar B_s^0\to\Lambda_c^+\bar\Lambda_c^-)
=(8.1^{+1.7}_{-1.5})\times10^{-5}$, which agrees with the experimental data within the quoted
uncertainties, although its central value is larger. A recent topological amplitude study further incorporated sizable $\text{SU}(3)$-breaking effects~\cite{Chua:2026awd}. 
Two-body charmed anti-charmed baryonic $B$ decays have also been studied in the perturbative-QCD approach with higher-twist corrections, where the importance of W-exchange topologies was also emphasized~\cite{Rui:2024xgc}. The decay $\bar B_s^0 \to \Lambda_c^+\bar{\Lambda}_c^-$ was predicted to exhibit a sizable direct CP asymmetry when only pseudoscalar-meson intermediate-state scatterings are considered within final-state interactions~\cite{Geng:2025yna}, which leads to an obviously larger branching fraction of order $\mathcal{O}(10^{-4})$. It indicates that some necessary contributions need to be included.

Recently, the final state interactions (FSIs) approach has been widely applied to a variety of non-leptonic decays of heavy-flavor hadrons and has shown promising phenomenological performance \cite{Jia:2024pyb,Feng:2026soj,Shang:2026knt,Duan:2024zjv,Hu:2025pjg,Hu:2024uia,Duan:2026oxg}. In this work, we investigate $\bar B^{0}_{(s)} \to \Lambda_c^+\bar{\Lambda}_c^-$ within the FSIs framework, taking into account both pseudoscalar- and vector-meson intermediate states. The hadronic loop integrals are regulated by introducing an effective form factor with the same cutoff parameters taken from our previous study of two-body nonleptonic decays of $\Lambda^{0}_{b}$~\cite{Duan:2024zjv}. These cutoff parameters were determined from experimental data, hence no additional free parameters are introduced in this work. Our results show good agreement with experimental measurements given in Eq.(1.1) and (1.2). Furthermore, we predict the direct CP asymmetries, the asymmetry parameters,
and the CP asymmetries associated with these parameters. We hope that these results provide new insight into the underlying dynamics of $\bar B^{0}_{(s)} \to \Lambda_c^+\bar{\Lambda}_c^-$ and serve as useful benchmarks for future experimental tests.

\section{Theoretical framework}\label{framework}

In this section, we introduce the final-state rescattering framework for
$\bar B^0_{(s)}\to\Lambda_c^+\bar\Lambda_c^-$. Following
Refs.~\cite{Jia:2024pyb,Duan:2024zjv}, we evaluate the long-distance FSI
amplitudes using the hadronic-loop approach. The weak decay vertices are
treated in naive factorization, whereas the strong-interaction vertices are
derived from Chiral Effective Lagrangians.

\subsection{Effective Hamiltonian}

The weak transitions are governed by the effective
Hamiltonian~\cite{Buchalla:1995vs},
\begin{equation}
  \mathcal{H}_{\rm eff}=\frac{G_F}{\sqrt{2}}
  \left\{\sum_{p=u,c}V_{pb}V^{*}_{pq}
  \Big[C_1(\mu)\,O^{p}_1(\mu)+C_2(\mu)\,O^{p}_2(\mu)\Big]
  -V_{tb}V^{*}_{tq}\sum_{i=3}^{10}C_i(\mu)\,O_i(\mu)\right\}+{\rm H.c.},
  \label{eq:Heff}
\end{equation}
where $q=d$ for $\bar B^{0}$ and $q=s$ for $\bar B^{0}_{s}$ decays. Here, $G_F$ is the Fermi constant, $V_{ij}$ are CKM matrix elements, and
$C_i(\mu)$ are the Wilson coefficients evaluated at the scale $\mu\simeq m_b$. The
current--current operators read
\begin{equation}
  O^{p}_1=(\bar p_\alpha b_\alpha)_{V-A}\,(\bar q_\beta p_\beta)_{V-A},
  \qquad
  O^{p}_2=(\bar p_\alpha b_\beta)_{V-A}\,(\bar q_\beta p_\alpha)_{V-A},
  \qquad p=u,c ,
\end{equation}
and the QCD penguin operators are
\begin{equation}
\begin{aligned}
  O_3&=(\bar q_\alpha b_\alpha)_{V-A}\sum_{q'}(\bar q'_\beta q'_\beta)_{V-A},
  &\qquad
  O_4&=(\bar q_\beta b_\alpha)_{V-A}\sum_{q'}(\bar q'_\alpha q'_\beta)_{V-A}, \\
  O_5&=(\bar q_\alpha b_\alpha)_{V-A}\sum_{q'}(\bar q'_\beta q'_\beta)_{V+A},
  &\qquad
  O_6&=(\bar q_\beta b_\alpha)_{V-A}\sum_{q'}(\bar q'_\alpha q'_\beta)_{V+A},
\end{aligned}
\end{equation}
where $\alpha$ and $\beta$ are color indices, $q'=u,d,s,c,b$, and $(\bar q_1 q_2)_{V\mp A}\equiv\bar q_1\gamma_\mu(1\mp\gamma_5)q_2$. The electroweak penguin operators $O_7$--$O_{10}$ are defined as
in Ref.~\cite{Duan:2024zjv}. Their contributions are neglected because
the corresponding Wilson coefficients are numerically small. After
factorizing the four-quark matrix elements, the color contractions yield
\begin{equation}
  a_i=C_i+\frac{C_{i+1}}{N_c}\quad(i=1,3,5),
  \qquad
  a_i=C_i+\frac{C_{i-1}}{N_c}\quad(i=2,4,6),
  \label{eq:ai}
\end{equation}
where $N_c=3$. The Wilson coefficients at $\mu=m_b$ are taken from
Ref.~\cite{Lu:2009cm}.

\subsection{Final-state rescattering mechanism and long-distance amplitudes}

Following Refs.~\cite{Jia:2024pyb,Duan:2024zjv}, we describe final-state
interactions using the hadron-level time-evolution picture. The weak decay
occurs on a short timescale $\tau$, after which the produced hadrons undergo
long-distance strong interactions. For a final state $f$, the decay amplitude
can be written as
\begin{equation}
  \mathcal A(\bar B^0_{(s)}\to f)
  =\sum_i
  \langle f|U(+\infty,\tau)|i\rangle
  \langle i|\mathcal H_{\rm eff}|\bar B^0_{(s)}\rangle ,
  \label{eq:fsi_master}
\end{equation}
where the sum runs over all possible hadronic states $|i\rangle$. The matrix element $\langle i|\mathcal H_{\rm eff}|\bar B^0_{(s)}\rangle$ is dominated by the short-distance dynamics of the weak interaction scale and is evaluated in naive
factorization. Subsequently, an intermediate state $i$ transits to the final state $f$ through long distant strong interaction, which is described by the matrix element $\langle f|U(+\infty,\tau)|i\rangle$. Strong phases, which are essential for evaluation of the direct CP violations, are generated in the long distance interactions.

\begin{figure}[H]
\centering
\includegraphics[width=0.92\textwidth]{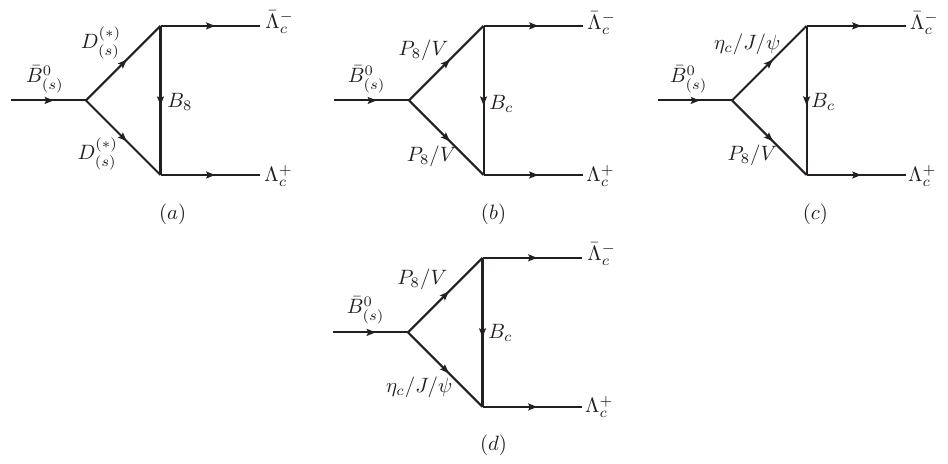}
\caption{
Triangle diagrams for
$\bar B^0_{(s)}\to \Lambda_c^+\bar\Lambda_c^-$
within the final-state re-scattering mechanism. Here $B_c$ and $B_8$ denote charmed baryons and the light-baryon octet, respectively, while $P_8$ and $V$ represent the pseudoscalar-meson octet and vector-meson octet.
}
\label{fig:Bmeson_triangle}
\end{figure}

A complete calculation of Eq.~\eqref{eq:fsi_master} should consider all possible intermediate states, which is impossible to achieve. We therefore follow the traditional opinion of FSI calculations and assumes that the dominant contributions arise from $2\to 2$ rescattering by single-particle exchange in the $t$ or $u$ channel.
The resulting long-distance contributions are described by triangle
diagrams. Among all the intermediate states, ground states are expected to contribute dominantly. In this paper, we retain ground-state light pseudoscalar and vector mesons, charmed
mesons and charmonia, the light-baryon octet, and the ground-state
charmed-baryon multiplets $\mathcal B_{\bar 3}$ and $\mathcal B_6$. The total
rescattering amplitude is obtained by coherently summing all allowed triangle
amplitudes listed in Appendix~\ref{app.A}. Representative diagrams for these rescattering processes are shown in Fig.~\ref{fig:Bmeson_triangle}.

As an explicit example, consider
\begin{equation*}
  \bar B^0(p_i)\to D^-(p_1)D^+(p_2)
  \to\Lambda_c^+(p_4,\lambda_4)\bar\Lambda_c^-(p_3,\lambda_3),
\end{equation*}
where $p_{1,2}$ and $p_{3,4}$ denote the intermediate-meson and final-baryon
momenta, respectively, and $k$ is the exchanged-neutron momentum. The $D^-$
is emitted from the weak current, while the $D^+$ contains the spectator
antiquark. In naive factorization, the weak matrix element is
\begin{equation}
\begin{aligned}
  &\langle D^-(p_1)D^+(p_2)|\mathcal H_{\rm eff}|
  \bar B^0(p_i)\rangle\\
  &=Z_{D^-}
  \langle D^-(p_1)|\bar d\gamma_\mu(1-\gamma_5)c|0\rangle
  \langle D^+(p_2)|\bar c\gamma^\mu(1-\gamma_5)b|
  \bar B^0(p_i)\rangle ,
  \label{eq:weak_DD_factorized}
\end{aligned}
\end{equation}
where$Z_{D^-}=\frac{G_F}{\sqrt{2}}
\Bigl[V_{cb}V_{cd}^{*}a_1
-V_{tb}V_{td}^{*}\bigl(a_4+r_\chi^{D^-}a_6\bigr)\Bigr]$,
with $r_\chi^{D^-}=2m_{D^-}^2/[(m_b-m_c)(m_c+m_d)]$
denoting the corresponding chiral factor.
The decay constant and the $\bar B^0\to D^+$ form factors are defined by
\begin{equation}
  \langle D^-(p_1)|\bar d\gamma_\mu(1-\gamma_5)c|0\rangle
  =i f_{D^-}p_{1\mu},
  \label{eq:fD_definition}
\end{equation}
and
\begin{equation}
\begin{aligned}
  \langle D^+(p_2)|\bar c\gamma^\mu b|\bar B^0(p_i)\rangle
  ={}&F_1^{\bar B^0\to D^+}(q^2)
  \left[(p_i+p_2)^\mu
  -\frac{M_{\bar B^0}^2-m_{D^+}^2}{q^2}q^\mu\right]+F_0^{\bar B^0\to D^+}(q^2)
  \frac{M_{\bar B^0}^2-m_{D^+}^2}{q^2}q^\mu,
  \label{eq:BD_form_factor_definition}
\end{aligned}
\end{equation}
respectively, where $q=p_i-p_2=p_1$. The axial-current matrix element between the two pseudoscalar mesons vanishes. Therefore, Eq. \eqref{eq:weak_DD_factorized} can be further evaluated as
\begin{equation}
  \langle D^-(p_1)D^+(p_2)|\mathcal H_{\rm eff}|
  \bar B^0(p_i)\rangle
  =Z_{D^-}\,i f_{D^-}\big(M_{\bar B^0}^2-m_{D^+}^2\big)
  F_0^{\bar B^0\to D^+}\big(m_{D^-}^2\big).
  \label{eq:weak_DD_example}
\end{equation}

The strong vertices are assigned by the Feynman rules of type $PBB$ in Appendix B of
Ref.~\cite{Duan:2024zjv}, which are derived from
\begin{equation}
  \mathcal L_{\Lambda_cND}
  =g_{\Lambda_cND}\big(\bar\Lambda_c i\gamma_5DN+{\rm H.c.}\big).
  \label{eq:L_LcND_example}
\end{equation}
The corresponding matrix elements are
\begin{equation}
\begin{aligned}
  \langle\Lambda_c^+(p_4,\lambda_4)|i\mathcal L_{\Lambda_cND}|
  D^+(p_2)n(k,s)\rangle
  &=g_1\,\bar u(p_4,\lambda_4)i\gamma_5u_n(k,s),\\[3pt]
  \langle\bar\Lambda_c^-(p_3,\lambda_3)n(k,s)|
  i\mathcal L_{\Lambda_cND}|D^-(p_1)\rangle
  &=g_2\,\bar u_n(k,s)i\gamma_5v(p_3,\lambda_3).
  \label{eq:strong_DDn_vertices}
\end{aligned}
\end{equation}
Here, $g_1$ and $g_2$ denote the coupling constants at the two strong
vertices, respectively. The amplitude of the triangle
diagram can be derived as 
\begin{equation}
\begin{aligned}
  \mathcal M(D^-,D^+;n)={}&
  \int\!\frac{d^4k}{(2\pi)^4}
  \Big[Z_{D^-}\,i f_{D^-}
  \big(M_{\bar B^0}^2-m_{D^+}^2\big)
  F_0^{\bar B^0\to D^+}\big(m_{D^-}^2\big)\Big]g_1g_2 \\
  &\times
  \bar u(p_4,\lambda_4)\gamma_5(\not\!k+m_n)
  \gamma_5v(p_3,\lambda_3)\\
  &\times
  \frac{i^3\mathcal F(k^2,\Lambda_{\rm charm})}
  {(p_1^2-m_{D^-}^2+i\varepsilon)
   (p_2^2-m_{D^+}^2+i\varepsilon)
   (k^2-m_n^2+i\varepsilon)},
  \label{eq:DDn_example}
\end{aligned}
\end{equation}
where $\sum_su_n(k,s)\bar u_n(k,s)=\not\!k+m_n$ is used. The amplitudes for $PP$, $PV$, $VP$, and $VV$ modes in Appendix~\ref{app.B} are obtained in the same way. We evaluate the four-dimensional loop integral directly, rather than keeping
only its absorptive part through the optical theorem~\cite{Cheng:2004ru}.
Both dispersive and absorptive contributions are therefore included. 

The strong couplings are defined for on-shell hadrons, whereas the baryons forming a loop are generally off shell. Each loop is therefore multiplied by the form factor in Ref.~\cite{Duan:2024zjv},
\begin{equation}
  \mathcal F(k^2,\Lambda_i)
  =\frac{\Lambda_i^4}
  {(k^2-m_{\rm ex}^2)^2+\Lambda_i^4}.
  \label{eq:form_factor}
\end{equation}
It is normalized to unity at $k^2=m_{\rm ex}^2$, suppresses highly off-shell
contributions, and regulates the ultraviolet behavior of the integral. The regulator introduces no physical pole and hence generates no additional absorptive part.

As shown in Fig.~\ref{fig:Bmeson_triangle}, the two-meson intermediate
states that rescatter into $\Lambda_c^+\bar\Lambda_c^-$ fall into two
categories: charm-containing intermediate states, such as
$D_{(s)}^{(*)-}D_{(s)}^{(*)+}$ and $\eta_c(J/\psi)P(V)$, and
light-meson intermediate states. Following our previous work~\cite{Duan:2024zjv}, we assign separate
cutoff parameters to the two types of rescattering processes to
phenomenologically account for their different off-shell behavior and
hadronic structures:
\begin{equation}
  \Lambda_{\rm charm}=1.0\pm0.1~{\rm GeV},
  \qquad
  \Lambda_{\rm charmless}=0.5\pm0.1~{\rm GeV}.
\end{equation}
These values were determined in Ref.~\cite{Duan:2024zjv} from the
measured branching fractions and direct CP asymmetries of
$\Lambda_b^0\to p\pi^-$ and $pK^-$ and are used directly in the
present calculation. Thus, no additional free parameters are
introduced.

\section{Numerical results and discussions}
This section includes the input parameters, numerical results and some phenomenological discussions. We calculate the helicity amplitudes, branching fractions, direct CP asymmetries, and asymmetry parameters.
\subsection{Input parameters}
The masses of the relevant baryons, mesons, and quarks are
from the Particle Data Group~\cite{ParticleDataGroup:2024cfk}.
The $B_{(s)}\to M$ transition form factors included in the weak vertices
are from the following references:
Ref.~\cite{FermilabLattice:2015ilb} for $B^0\to D$ form factors,
Ref.~\cite{Martinelli:2024bov} for $B^0\to D^*$ form factors,
Ref.~\cite{Biswas:2022yvh} for $B^0\to\pi$ form factors,
Ref.~\cite{Flynn:2023qmi} for $B_s^0\to K$ form factors,
Ref.~\cite{Ciuchini:2026hxy} for $B^0\to K$ form factors,
Ref.~\cite{Horgan:2015vla} for $B^0\to K^*$ and $B_s^0\to K^*$ form factors,
Ref.~\cite{Gubernari:2018wyi} for $B^0\to\rho$ form factors,
Ref.~\cite{Melic:2025uha} for $B^0\to\eta$ and $B_s^0\to\eta$ form factors,
Ref.~\cite{Harrison:2023dzh} for $B_s^0\to D_s^*$ form factors,
Ref.~\cite{Gao:2019lta} for $B^0\to\omega$ form factors,
and Ref.~\cite{McLean:2019qcx} for $B_s^0\to D_s$ form factors.
The adopted form-factor inputs incorporate both established results
and recent developments, including lattice-QCD calculations, global
analyses of the available lattice data, and other nonperturbative
approaches.

The relevant decay constants and strong coupling constants for the
$\mathcal B_c \mathcal B_c P_8(V)$ and $\mathcal B_c \mathcal B_8 D^{(*)}$ vertices are
from Refs.~\cite{Duan:2024zjv,Yu:2017zst}, while the couplings of $\eta_c$
and $J/\psi$ to charmed baryons are adopted
from Ref.~\cite{Shen:2019evi}.

\subsection{Numerical results}

Using the parameters
$\Lambda_{\rm charm}=1.0\pm0.1~\mathrm{GeV}$ and
$\Lambda_{\rm charmless}=0.5\pm0.1~\mathrm{GeV}$
\cite{Duan:2024zjv}, we evaluate the two independent helicity amplitudes
$H_{++}$ and $H_{--}$ for
$\bar B^0_{(s)}\to\Lambda_c^+\bar\Lambda_c^-$ and decompose them into
CKM components as follows:
\begin{equation}
H_h(\bar B_q^0)
=
\sum_{p=u,c,t}
\lambda_p^{(q)}\,\mathcal H_{h,p}^{(q)},
\qquad
\lambda_p^{(q)}\equiv V_{pb}V_{pq}^{*},\qquad q=d,s .
\label{decomposion}
\end{equation}
Here, $h=++$ or $--$ labels the helicity configuration of the
final-state $\Lambda_c^+\bar\Lambda_c^-$ pair, and
$\mathcal H_{h,p}^{(q)}$ are the reduced helicity amplitudes without
CKM pre-factors. They contain the strong phases generated by final-state
rescattering, and their numerical values are listed in
Table~\ref{Helicity_B_to_LcLc}.

One can construct the branching fraction, the direct CP asymmetry, and the polarization observables in terms of helicity amplitudes. The decay width is given by
\begin{equation}
\Gamma(B\to \mathcal{B}\,\bar{\mathcal{B}})
=
\frac{|\textbf{p}|}{8\pi m_B^2}
\left(
|H_{++}|^2+|H_{--}|^2
\right),
\end{equation}
where \(m_B\) is the mass of \(B\) meson, and \(|\textbf{p}|\) denotes the magnitude of the $3$-momentum of the final-state baryon at the \(B\)-meson rest frame.

The direct CP asymmetry is
\begin{equation}
A_{CP}^{\rm dir}
=
\frac{
\left(|H_{++}|^2+|H_{--}|^2\right)
-
\left(|\bar H_{++}|^2+|\bar H_{--}|^2\right)
}{
\left(|H_{++}|^2+|H_{--}|^2\right)
+
\left(|\bar H_{++}|^2+|\bar H_{--}|^2\right)
},
\end{equation}
where $\bar H_{\pm\pm}$ denote the helicity amplitudes of the CP-conjugated decays. In the present framework, nonzero direct CP asymmetries arise from the interference between $\lambda_c\,\mathcal{H}_\lambda^{(c)}$ and the other two terms in Eq.\eqref{decomposion}, with strong phases in the reduced helicity amplitudes generated by final-state re-scattering effects. The numerical results of branching
ratios and direct CP asymmetries are listed in Table~\ref{BR_CP_B_to_LcLc}.

We also consider the asymmetry parameters defined as ~\cite{Rui:2024xgc}
\begin{equation}
\alpha=
\frac{|H_{++}|^2-|H_{--}|^2}{|H_{++}|^2+|H_{--}|^2},\qquad
\beta=
\frac{2\,{\rm Im}(H_{++}H_{--}^*)}{|H_{++}|^2+|H_{--}|^2},\qquad
\gamma=
\frac{2\,{\rm Re}(H_{++}H_{--}^*)}{|H_{++}|^2+|H_{--}|^2}.
\end{equation}
Here, $\alpha$ denotes the up-down asymmetry. Both $\alpha$ and
$\beta$ are $P$-odd, with $\beta$ also naively $T$-odd, whereas
$\gamma$ is $P$-even. These parameters satisfy the normalization condition $\alpha^2+\beta^2+\gamma^2=1$. The corresponding CP-violating observables associated with these asymmetry parameters can be defined as ~\cite{Donoghue:1986hh}
\begin{equation}
a_{CP}^{\alpha}
=
\frac{\alpha+\bar\alpha}{2},
\qquad
a_{CP}^{\beta}
=
\frac{\beta+\bar\beta}{2},
\qquad
a_{CP}^{\gamma}
=
\frac{\gamma-\bar\gamma}{2},
\end{equation}
where $\bar\alpha$, $\bar\beta$, and $\bar\gamma$ are the corresponding
asymmetry parameters for the CP-conjugate decay.

The partial-wave amplitudes are given by
\begin{equation}
S=\frac{H_{++}+ H_{--}}{\sqrt2},
\qquad
P=\frac{ H_{--}-  H_{++}}{\sqrt2}.
\end{equation}
We define the partial-wave CP asymmetries~\cite{Roy:2019cky}
\begin{equation}
a_{CP}^{S}=
\frac{|S|^{2}-|\bar S|^{2}}{|S|^{2}+|\bar S|^{2}},
\qquad
a_{CP}^{P}=
\frac{|P|^{2}-|\bar P|^{2}}{|P|^{2}+|\bar P|^{2}} .
\end{equation}

The numerical results for the asymmetry parameters $\alpha$, $\beta$, and
$\gamma$, the CP asymmetries associated with these parameters,
$a_{CP}^{\alpha}$, $a_{CP}^{\beta}$, and $a_{CP}^{\gamma}$, as well as
the partial-wave CP asymmetries, are listed in
Table~\ref{tab:asymmetry_parameters}.

\begin{table*}[h]
\centering
\caption{ Reduced helicity amplitudes (in units of $10^{-8}~\mathrm{GeV}$) for $\bar{B}_{(s)}^0 \to \Lambda_c^+\bar{\Lambda}_c^-$ with different CKM factors. Here $q=d$ for $\bar{B}^0$ and $q=s$ for $\bar{B}_s^0$.}
\renewcommand{\arraystretch}{1.35}
\setlength{\tabcolsep}{6pt}
\small
\begin{tabular}{c c c c}
\toprule
\toprule
CKM factor
& Helicity
& $\bar{B}^0\to \Lambda_c^+\bar{\Lambda}_c^-$
& $\bar{B}_s^0\to \Lambda_c^+\bar{\Lambda}_c^-$ \\
\midrule

\multirow{2}{*}{$V_{ub}V_{uq}^*$}
& $\mathcal H_{--,u}^{(q)}$
&
$3.68^{+3.94}_{-2.17}-1.51^{+0.89}_{-1.62}i$
&
$7.30^{+7.82}_{-4.31}-4.55_{-4.89}^{+2.69}i$
\\
& $\mathcal H_{++,u}^{(q)}$
&
$-0.62^{+0.37}_{-0.67}+3.34^{+3.58}_{-1.97}i$
&
$-6.51^{+3.84}_{-6.97}+4.77^{+5.12}_{-2.82}i$
\\
\midrule

\multirow{2}{*}{$V_{cb}V_{cq}^*$}
& $\mathcal H_{--,c}^{(q)}$
&
$264.52^{+109.90}_{-86.30}+94.51^{+31.77}_{-28.49}i$
&
$148.81^{+63.99}_{-49.44}+33.16^{+12.28}_{-10.44}i$
\\
& $\mathcal H_{++,c}^{(q)}$
&
$294.31^{+77.06}_{-74.78}-72.76^{+32.70}_{-44.77}i$
&
$122.90^{+34.61}_{-32.94}-89.09^{+29.46}_{-37.03}i$
\\
\midrule

\multirow{2}{*}{$V_{tb}V_{tq}^*$}
& $\mathcal H_{--,t}^{(q)}$
&
$-5.18^{+1.71}_{-1.39}+0.26^{+1.55}_{-1.20}i$
&
$-2.35^{+0.98}_{-0.79}+0.75^{+0.83}_{-0.70}i$
\\
& $\mathcal H_{++,t}^{(q)}$
&
$4.09^{+0.72}_{-1.06}-1.63^{+1.48}_{-1.61}i$
&
$1.83^{+0.60}_{-0.81}-1.72^{+0.89}_{-0.99}i$
\\

\bottomrule
\bottomrule
\end{tabular}
\label{Helicity_B_to_LcLc}
\end{table*}

\begin{table}[H]
\centering
\caption{Branching ratios and direct CP asymmetries}
\renewcommand{\arraystretch}{1.2}
\setlength{\tabcolsep}{5pt}
\begin{tabular}{>{\centering\arraybackslash}p{3.6cm}
                >{\centering\arraybackslash}p{1.8cm}
                >{\centering\arraybackslash}p{5.5cm}
                >{\centering\arraybackslash}p{3.0cm}
               }
\toprule
\toprule
Decay mode & Source & $\mathcal{BR}$ & $A_{\rm CP}^{\rm dir}$ \\
\midrule

\multirow{2}{*}{$\bar{B}^0 \to \Lambda_c^+ \bar{\Lambda}_c^-$}
& FSI
& $0.66^{+0.53}_{-0.33}\times 10^{-5}$
& $0.84^{+1.02}_{-2.59}\times10^{-3}$ \\

& Experiment
& $\left(1.01^{+0.27}_{-0.28}\pm 0.08 \pm 0.15\right)\times 10^{-5}$~\cite{LHCb:2025ueu}
& --- \\

\midrule

\multirow{2}{*}{$\bar{B}_s^0 \to \Lambda_c^+ \bar{\Lambda}_c^-$}
& FSI
& $3.43^{+3.10}_{-1.80}\times 10^{-5}$
& $-0.74^{+0.50}_{-1.30}\times 10^{-3}$ \\

& Experiment
& $(5.0 \pm 1.3 \pm 0.5 \pm 0.8)\times 10^{-5}$~\cite{LHCb:2025ueu}
& ---  \\
\bottomrule
\bottomrule
\end{tabular}
\label{BR_CP_B_to_LcLc}
\end{table}

\begin{table}[htbp]
\centering
\caption{
Asymmetry parameters, their CP asymmetries, and partial-wave CP asymmetries
}
\label{tab:asymmetry_parameters}
\renewcommand{\arraystretch}{1.2}
\setlength{\tabcolsep}{3pt}
\small
\begin{tabular}{cccccccccc}
\toprule
\toprule
Decay mode & $\alpha$ & $\beta$ & $\gamma$ & $a_{CP}^{\alpha}$ & $a_{CP}^{\beta}$  & $a_{CP}^{\gamma}$& $a_{CP}^S$ & $a_{CP}^P$\\
\hline
$\bar{B}^0 \to \Lambda_c^+\bar{\Lambda}_c^-$
& $0.04_{-0.09}^{+0.08}$  & $-0.56^{+0.05}_{-0.04}$ & $0.83^{+0.02}_{-0.03}$ & $0.05^{+0.09}_{-0.09}$ & $-0.55^{+0.05}_{-0.04}$ & $-0.008^{+0.001}_{-0.003}$ & $-4^{+2}_{-4}\times10^{-3}$ & $0.05^{+0.02}_{-0.01}$\\

$\bar{B}_s^0 \to \Lambda_c^+\bar{\Lambda}_c^-$
& $-0.04_{-0.07}^{+0.07}$  & $-0.74^{+0.02}_{-0.02}$ & $0.67^{+0.03}_{-0.03}$ & $-0.04^{+0.07}_{-0.07}$  & $-0.74^{+0.02}_{-0.02}$ & $0.001^{+0.002}_{-0.001}$& $5^{+9}_{-3}\times10^{-5}$ & $-5^{+3}_{-9}\times 10^{-3}$ \\
\toprule
\toprule
\end{tabular}
\end{table}

\subsection{Discussion}
 Based on the above numerical results, several essential discussions are in order:

 \begin{itemize}

\item In this work, we take
$\Lambda_{\rm charm}=1.0\pm0.1~\mathrm{GeV}$ and
$\Lambda_{\rm charmless}=0.5\pm0.1~\mathrm{GeV}$, with the central
predictions evaluated at $1.0~\mathrm{GeV}$ and $0.5~\mathrm{GeV}$,
respectively. We scan the two parameters independently over their respective
ranges, with
no correlation imposed, and assign the largest positive and negative deviations from the
central value of each observable as its upper and lower uncertainties.
The quoted uncertainties therefore reflect only the cutoff-induced model
dependence and do not correspond to a specific statistical confidence level.

The weak-transition form factors constitute another important theoretical
input. Since the long-distance contributions are dominated by rescattering
through charmed-meson intermediate states, the relevant
$B_{(s)}\to D_{(s)}^{(*)}$ form factors are evaluated at
$q^2\simeq m_{D_{(s)}^{(*)}}^2$. For a reliable description of this
kinematic region, we use direct lattice-QCD results or global analyses
constrained by lattice data. For comparison, we repeat
the $\bar B^0$ calculation using the LCSR form factors of
Ref.~\cite{Gubernari:2018wyi}, keeping all other inputs unchanged. The
branching fraction changes only from $0.66\times10^{-5}$ to
$0.67\times10^{-5}$, while the direct $CP$ asymmetry remains of order
$10^{-3}$. The main effect is a redistribution of the two dominant helicity
amplitudes rather than a change in their summed strength. Consequently,
$\alpha$ changes from $0.04$ to $-0.58$, whereas the changes in $\beta$ and
$\gamma$ are more moderate. Thus, for the two form-factor choices considered here, the branching fraction and direct $CP$ asymmetry remain stable, whereas helicity-sensitive observables, particularly $\alpha$, exhibit a stronger form factor dependence.

\item
As shown in Table~\ref{Helicity_B_to_LcLc}, the amplitude component associated with the CKM factor $V_{cb}V_{cq}^{*}$ is dominant for both helicity amplitudes in both decay modes.  Thus, these
decays are driven mainly by the $b\to c\bar c q$ tree transition. The
components associated with $V_{ub}V_{uq}^{*}$ and $V_{tb}V_{tq}^{*}$
are much smaller. The suppression of the $V_{tb}V_{tq}^{*}$ component is consistent with the fact that $a_1$ is typically one to two orders of magnitude larger than $a_4+r_\chi a_6$.

\item 
 We employ the same cutoff parameters as those used in Ref.~\cite{Duan:2024zjv}, and obtain branching fractions for $\bar B^0\to \Lambda_c^+\bar\Lambda_c^-$ and $\bar B_s^0\to \Lambda_c^+\bar\Lambda_c^-$ that are compatible with the updated experimental measurements. In Ref.~\cite{Duan:2024zjv}, the model parameters $\Lambda_{\text{charm}}$ and $\Lambda_{\text{charmless}}$ were determined applying the measured branching fractions and CP asymmetries of the decays $\Lambda^{0}_{b}\to p\pi^{-},pK^{-}$.  It is therefore natural to expect that these parameters should also be applicable to $B$ meson decays, since both $\Lambda_b$ and $B$ decays are weak processes governed by the effective interaction at the $b$ quark mass scale. As a result, the associated long-distance final-state re-scattering effects in these decays are expected to be analogous. Meanwhile, our numerical results further support this expectation.

\item 
In Table~\ref{BR_CP_B_to_LcLc}, the direct CP asymmetries predicted in this work for decays $\bar B^0$ and $\bar B_s^0\to \Lambda_c^+\bar\Lambda_c^-$
are found to be close to zero. It is in clear different from the $-10\%$ direct CP asymmetry predicted for $\bar B_s^0\to \Lambda_c^+\bar\Lambda_c^-$ in Ref.~\cite{Geng:2025yna}. The difference originates from the treatment of final state interactions, where Ref.~\cite{Geng:2025yna} includes only pseudoscalar intermediate states in the re‑scattering amplitudes, while the present analysis incorporates both pseudoscalar and vector intermediate states, namely the $PP$, $PV$, $VP$, and $VV$ contributions. Since the branching fractions of the $B\to D^*D^*$ modes are significantly larger than those of the corresponding $B\to DD$ modes, intermediate states containing charmed vector mesons $D_{(s)}^*$ can substantially enhance the long-distance amplitudes proportional to $V_{cb}V_{cq}^{*}$ $(q=d,s)$. Consequently, the interference between amplitudes with different weak phases is strongly suppressed, resulting in nearly vanishing direct CP asymmetries.  Meanwhile, Table~\ref{tab:asymmetry_parameters} shows that the
partial-wave CP asymmetries are generally small. The $S$-wave
asymmetries are below the percent level in both decay modes, while
the $P$-wave asymmetry reaches approximately $5\%$ for
$\bar B^0\to\Lambda_c^+\bar\Lambda_c^-$ and remains below the
percent level for
$\bar B_s^0\to\Lambda_c^+\bar\Lambda_c^-$. The nearly vanishing
total direct CP asymmetries result from the dominance of the tree
amplitudes, together with the relative weights and partial cancellation
of the $S$- and $P$-wave contributions.

\item
Table~\ref{tab:asymmetry_parameters} summarizes our predictions for the
asymmetry parameters. The parameter $\alpha$, which characterizes the
longitudinal polarization of the final-state $\Lambda_c^+$, is close to
zero in both decay modes. Our result for the $\bar B_s^0$ mode agrees
with Ref.~\cite{Rui:2024xgc}. For the $\bar B^0$ mode, that reference
gives a larger central value, although the quoted uncertainty intervals
overlap. Both calculations predict sizable negative values of $\beta$
in the two modes, with a larger numerical difference for the
$\bar B^0$ mode. Their central predictions for $\gamma$, however, have
opposite signs in both modes. These observables are sensitive to the
helicity structure: $\alpha$ mainly probes the relative magnitudes of
the two helicity amplitudes, whereas $\beta$ and $\gamma$ probe their
interference and relative phase. The differences may therefore reflect
the distinct strong-interaction dynamics: Ref.~\cite{Rui:2024xgc}
employs PQCD based on hard-gluon exchange, whereas the present
calculation considers long-distance final-state rescattering through
pseudoscalar and vector intermediate states.

 \end{itemize}

\section{Summary}
In this work, we study the branching fractions, direct CP asymmetries, and the asymmetry parameters $\alpha$, $\beta$, $\gamma$ of the charmed two-body decays
$\bar B^0_{(s)}\to \Lambda_c^+\bar\Lambda_c^-$ within the framework of final-state re-scatterings. Using the cutoff parameters determined from $\Lambda^{0}_{b}$ baryon decays, we find that the predicted branching fractions are consistent with current experimental measurements, indicating that long‑distance re‑scattering effects play an essential role in these charmed $B$ decays. 
In particular, these long-distance contributions provide a candidate dynamical explanation for observed enhancement of the W-exchange contribution that is largely suppressed in the naive factorization estimation. The direct CP asymmetries in both channels are found to be nearly vanishing as the inclusion of charmed vector intermediate states significantly enhances the tree contributions, thereby suppressing the interference between amplitudes with different weak phases. Therefore, measurements of CP asymmetries in these modes are crucial for determining whether vector meson intermediate re‑scattering effects are indeed indispensable. As for the polarization observables, both decay modes have values of $\alpha$ close to zero and hence exhibit suppressed longitudinal polarizations. In contrast, the sizable values of $\beta$ and $\gamma$ indicate substantial interference between the two helicity amplitudes. These predictions can be tested in future experimental studies, and the same methodology can be applied to investigate other $B$
meson decays into charmed baryon pairs.
\section*{Acknowledgments}

This work was supported in part by the China Postdoctoral Science Foundation under Grant No.~GZC20261850, by the National Natural Science Foundation of China under Grant Nos.~12335003, 12075126, and 12247101, by the Scientific Research Innovation Capability Support Project for Young Faculty under Grant No.~ZYGXQNJSKYCXNLZCXM-P2, by the Fundamental Research Funds for the Central Universities under Grant Nos.~lzujbky-2023-stlt01, lzujbky-2023-it12, lzujbky-2024-oy02, lzujbky-2025-eyt01, and lzujbky-2025-jdzx07, by the Natural Science Foundation of Gansu Province under Grant No.~25JRRA799, and by the `111 Center' under Grant No.~B20063.

\appendix
\section{Total Amplitudes for $\bar B_{(s)}^0\to \Lambda_c^+\bar{\Lambda}_c^-$}\label{app.A}

Here, we list the total amplitudes of $\bar{B}^0,\bar{B}_s^0\to \Lambda_c^+ \bar{\Lambda}_c^-$. The notation $\mathcal{M}(A,B;C)$ denotes the triangle diagram amplitude in which the initial $\bar{B}_{(s)}^0$ decays  into the intermediate states $A$ and $B$, followed by their re-scattering through the exchange of $C$.
\begin{equation}
    \begin{aligned}
        &\mathcal{A}(\bar{B}^0\to \Lambda_c^+ \bar{\Lambda}_c^-)\\
        &=\mathcal{M}(D^-,D^+;n)+\mathcal{M}(\pi^-,\pi^+;\Sigma_c^0)+\mathcal{M}(\pi^+,\pi^-;\Sigma_c^{++})+\mathcal{M}(\pi^0,\pi^0;\Sigma_c^+)+\mathcal{M}(\bar{K}^0,K^0;\Xi_c^+)+\mathcal{M}(\bar{K}^0,K^0;\Xi_c^{\prime +})\\
        &+\mathcal{M}(\eta,\eta;\Lambda_c^+)+\mathcal{M}(\eta_c,\eta;\Lambda_c^+)+\mathcal{M}(\eta,\eta_c;\Lambda_c^+)+\mathcal{M}(D^-,D^{*+};n)
+\mathcal{M}(\pi^+,\rho^-;\Sigma_c^{++})
+\mathcal{M}(\pi^-,\rho^+;\Sigma_c^0)\\
&+\mathcal{M}(\pi^0,\rho^0;\Sigma_c^+)
+\mathcal{M}(\bar{K}^0,K^{*0};\Xi_c^+)
+\mathcal{M}(\bar{K}^0,K^{*0};\Xi_c^{\prime +})
+\mathcal{M}(\eta,\omega;\Lambda_c^+)
+\mathcal{M}(\eta_c,\omega;\Lambda_c^+)
+\mathcal{M}(\eta,J/\psi;\Lambda_c^+)\\
&+\mathcal{M}(D^{*-},D^+;n)
+\mathcal{M}(\rho^-,\pi^+;\Sigma_c^0)
+\mathcal{M}(\rho^+,\pi^-;\Sigma_c^{++})
+\mathcal{M}(\rho^0,\pi^0;\Sigma_c^+)
+\mathcal{M}(\bar{K}^{*0},K^0;\Xi_c^+)+\mathcal{M}(\bar{K}^{*0},K^0;\Xi_c^{\prime +})\\
&+\mathcal{M}(\omega,\eta;\Lambda_c^+)
+\mathcal{M}(J/\psi,\eta;\Lambda_c^+)
+\mathcal{M}(\omega,\eta_c;\Lambda_c^+)
+\mathcal{M}(D^{*-},D^{*+};n)
+\mathcal{M}(\rho^-,\rho^+;\Sigma_c^0)
+\mathcal{M}(\rho^+,\rho^-;\Sigma_c^{++})\\
&+\mathcal{M}(\rho^0,\rho^0;\Sigma_c^+)
+\mathcal{M}(\bar{K}^{*0},K^{*0};\Xi_c^+)
+\mathcal{M}(\bar{K}^{*0},K^{*0};\Xi_c^{\prime +})
+\mathcal{M}(\omega,\omega;\Lambda_c^+)
+\mathcal{M}(\omega,J/\psi;\Lambda_c^+)
+\mathcal{M}(J/\psi,\omega;\Lambda_c^+),
    \end{aligned}
\end{equation}
\begin{equation}
    \begin{aligned}
       &\mathcal{A}(\bar{B}_s^0\to \Lambda_c^+ \bar{\Lambda}_c^-)\\
        &=\mathcal{M}(D_s^-,D_s^+;\Lambda^0)
+\mathcal{M}(K^-,K^+;\Xi_c^0)
+\mathcal{M}(K^-,K^+;\Xi_c^{\prime 0})
+\mathcal{M}(\bar{K}^0,K^0;\Xi_c^+)
+\mathcal{M}(\bar{K}^0,K^0;\Xi_c^{\prime +})
+\mathcal{M}(\eta,\eta;\Lambda_c^+)\\
&+\mathcal{M}(\eta,\eta_c;\Lambda_c^+)
+\mathcal{M}(\eta_c,\eta;\Lambda_c^+)+\mathcal{M}(D_s^-,D_s^{*+};\Lambda^0)
+\mathcal{M}(K^-,K^{*+};\Xi_c^0)
+\mathcal{M}(K^-,K^{*+};\Xi_c^{\prime 0})
+\mathcal{M}(\bar{K}^0,K^{*0};\Xi_c^+)\\
&+\mathcal{M}(\bar{K}^0,K^{*0};\Xi_c^{\prime +})
+\mathcal{M}(\eta,\omega;\Lambda_c^+)
+\mathcal{M}(\eta,J/\psi;\Lambda_c^+)
+\mathcal{M}(D_s^{*-},D_s^+;\Lambda^0)
+\mathcal{M}(K^{*-},K^+;\Xi_c^0)
+\mathcal{M}(K^{*-},K^+;\Xi_c^{\prime 0})\\
&+\mathcal{M}(\bar{K}^{*0},K^0;\Xi_c^{\prime +})
+\mathcal{M}(\bar{K}^{*0},K^0;\Xi_c^+)
+\mathcal{M}(\omega,\eta;\Lambda_c^+)
+\mathcal{M}(J/\psi,\eta;\Lambda_c^+)
+\mathcal{M}(D_s^{*-},D_s^{*+};\Lambda^0)
+\mathcal{M}(K^{*-},K^{*+};\Xi_c^0)\\
&+\mathcal{M}(K^{*-},K^{*+};\Xi_c^{\prime 0})
+\mathcal{M}(\bar{K}^{*0},K^{*0};\Xi_c^+)
+\mathcal{M}(\bar{K}^{*0},K^{*0};\Xi_c^{\prime +}).
    \end{aligned}
\end{equation}

\section{Amplitudes of triangle diagram}\label{app.B}
Below we present the unified expressions for the triangle-diagram amplitudes shown in Fig.~\ref{fig:Bmeson_triangle}. 
The relevant processes can be classified into four categories, in which the weak decay of $\bar B_{(s)}^0$ produces the intermediate two-meson states $PP$, $PV$, $VP$, and $VV$, followed by re-scattering through the exchange of a baryon $\mathcal B$. Here, the $PV$ and $VP$ intermediate states correspond to the $t$- and $u$-channel contributions, respectively. The symbols $P_{1,2}$ and $V_{1,2}$ denote pseudoscalar and vector mesons, respectively.
\begin{equation}
    \begin{aligned}
        &M[P_1,P_2;\mathcal B]\\
        &=\int \frac{d^4k}{(2\pi)^4}\Big(Z_1\,if_{P_2}(M_B^2-M_{P_1}^2)F_{0}^{B\to P_1}(M_{P_2}^2)+Z_2\,if_{P_1}(M_B^2-M_{P_2}^2)F_{0}^{B \to  P_2}(M_{P_1}^2)\Big)\\
        &\times g_{1PBB}\,g_{2PBB}\,\bar{u}(p_4,\lambda_4)\gamma_5(\not\!k+m_k)\gamma_5v(p_3,\lambda_3)\frac{{i^3}\mathcal{F}(k^2,\Lambda)}{(p_1^2-m_1^2+i\varepsilon)(p_2^2-m_2^2+i\varepsilon)(k^2-m_k^2+i\varepsilon)},
    \end{aligned}
\end{equation}
\begin{equation}
    \begin{aligned}
         &M[P,V;\mathcal B]\\
        &=\int \frac{d^4k}{(2\pi)^4}\Big(Z_1\, 2f_{V}M_{V}\,F_1^{B\to P}(M_V^2)+Z_2\,2f_{P}M_V A_{0}^{B \to V}(M_P^2) \Big)\,p_{i\alpha}\,(-g^{\alpha \mu}+\frac{p_2^\alpha p_2^\mu}{m_2^2}) \,(-ig_{1PBB}) \\
        &\times  \,\bar{u}(p_4,\lambda_4)\Big(f_{1VBB}\gamma_\mu-\frac{i\,f_{2VBB}}{m_k+m_4}\sigma_{\nu\mu}p_2^\nu \Big)(\not\!k +m_k
        )\gamma_5v(p_3,\lambda_3)\cdot \frac{{i^3}\mathcal{F}(k^2,\Lambda)}{(p_1^2-m_1^2+i\varepsilon)(p_2^2-m_2^2+i\varepsilon)(k^2-m_k^2+i\varepsilon)},\\
    \end{aligned}
\end{equation}
\begin{equation}
    \begin{aligned}
         &M[V,P;\mathcal B]\\
        &=\int \frac{d^4k}{(2\pi)^4}\Big(Z_1\,2f_{P}M_VA_{0}^{B \to V}(M_P^2) +Z_2\, 2f_{V}M_{V}\,F_1^{B\to P}(M_V^2)\Big)p_{i\alpha}(-g^{\alpha \mu}+\frac{p_1^\alpha p_1^\mu}{m_1^2})\cdot (-ig_{PBB})\\
        &\times \bar{u}(p_4,\lambda_4)\gamma_5(\not\!k + m_k)\Big(f_{1VBB}\gamma_\mu-\frac{i\,f_{2VBB}}{m_k+m_3}\sigma_{\nu\mu}p_1^\nu \Big)v(p_3,\lambda_3)\frac{{i^3}\mathcal{F}(k^2,\Lambda)}{(p_1^2-m_1^2+i\varepsilon)(p_2^2-m_2^2+i\varepsilon)(k^2-m_k^2+i\varepsilon)},
    \end{aligned}
\end{equation}
\begin{equation}
    \begin{aligned}
         &M[V_1,V_2;\mathcal B]\\
        &=\int \frac{d^4k}{(2\pi)^4}\Bigg\{
        Z_1\,
        f_{V_2}M_{V_2}\Bigg[ \frac{2V^{B\to V_1}(M_{V_2}^2)}{M_{B}+M_{V_1}}\epsilon_{\mu\nu\alpha\beta}\;(-g^{\mu\delta}+\frac{p_2^\mu p_2^\delta}{m_2^2})(-g^{\nu\rho}+\frac{p_1^\nu p_1^\rho}{m_1^2})p_{i}^{\alpha}p_{1}^{\beta} -i\cdot(M_{B}+M_{V_1})\\
        &\times(-g_{\mu\delta}+\frac{p_{2\mu} p_{2\delta}}{m_2^2})(-g^{\mu\rho}+\frac{p_1^\mu p_1^\rho}{m_1^2})A_{1}^{B\to V_1}(M_{V_2}^2)+i\cdot\frac{2p_{i\mu} p_{i\nu}}{M_{B}+M_{V_1}} (-g^{\mu\delta}+\frac{p_2^\mu p_2^\delta}{m_2^2})(-g^{\nu\rho}+\frac{p_1^\nu p_1^\rho}{m_1^2})A_2^{B\to V_1}(M_{V_2}^2)\Bigg]\\
        &+Z_2\,f_{V_1}M_{V_1}\Bigg[ \frac{2V^{B\to V_2}(M_{V_1}^2)}{M_{B}+M_{V_2}}\epsilon_{\mu\nu\alpha\beta}\;(-g^{\nu\delta}+\frac{p_2^\nu p_2^\delta}{m_2^2})(-g^{\mu\rho}+\frac{p_1^\mu p_1^\rho}{m_1^2})p_{i}^{\alpha}p_{2}^{\beta}-i\cdot(M_{B}+M_{V_2})(-g^{\mu\delta}+\frac{p_2^\mu p_2^\delta}{m_2^2})\\
        &\times(-g_{\mu\rho}+\frac{p_{1\mu} p_{1\rho}}{m_1^2})A_{1}^{B\to V_2}(M_{V_1}^2)+i\cdot\frac{2 p_{i\mu} p_{i\nu}}{M_{B}+M_{V_2}}(-g^{\mu\delta}+\frac{p_2^\mu p_2^\delta}{m_2^2})(-g^{\nu\rho}+\frac{p_1^\nu p_1^\rho}{m_1^2}) A_2^{B\to V_2}(M_{V_1}^2)\Bigg]\Bigg\}\cdot  (-1)\,\\
        &\times \bar{u}(p_4,\lambda_4)\Big(f_{1VBB}\gamma_\delta-\frac{i\,f_{2VBB}}{m_k+m_4}\sigma_{\tau\delta}p_2^\tau \Big)(\not\!k +m_k)\Big(f_{1VBB}\gamma_\rho-\frac{i\,f_{2VBB}}{m_k+m_3}\sigma_{\gamma\rho}p_1^\gamma \Big)v(p_3,\lambda_3)\\
        &\times \frac{{i^3}\mathcal{F}(k^2,\Lambda)}{(p_1^2-m_1^2+i\varepsilon)(p_2^2-m_2^2+i\varepsilon)(k^2-m_k^2+i\varepsilon)}.\\
    \end{aligned}
\end{equation}

In the naive-factorization treatment of the weak transition
$B\to M_1M_2$, the two possible factorizable contributions are weighted
by the channel-dependent coefficients $Z_{1,2}$, which contain the
relevant CKM factors and the corresponding combinations of effective
Wilson coefficients, as illustrated by $Z_{D^-}$ in Sec.~2.2.

The pseudoscalar- and vector-meson decay constants are defined by
\begin{equation}
\begin{aligned}
  \langle P(p)|\bar q_1\gamma_\mu(1-\gamma_5)q_2|0\rangle
  &=i f_P p_\mu,\\
  \langle V(p,\epsilon)|\bar q_1\gamma_\mu(1-\gamma_5)q_2|0\rangle
  &=m_V f_V\epsilon_\mu^*.
\end{aligned}
\label{eq:decay_constant_definitions}
\end{equation}
The $\bar B_{(s)}\to P$ transition form factors are defined by~\cite{Gubernari:2018wyi}
\begin{equation}
\begin{aligned}
  &\langle P(p_P)|\bar q\gamma_\mu b|\bar B(p_B)\rangle\\
  &\quad =
  F_1^{B\to P}(q^2)
  \left[
    (p_B+p_P)_\mu
    -\frac{m_B^2-m_P^2}{q^2}q_\mu
  \right]
  +F_0^{B\to P}(q^2)
  \frac{m_B^2-m_P^2}{q^2}q_\mu,
\end{aligned}
\label{eq:BtoP_form_factor_definition}
\end{equation}
where $q=p_B-p_P$ and $F_1^{B\to P}(0)=F_0^{B\to P}(0)$.
For the $\bar B_{(s)}\to V$ transitions, we use~\cite{Gubernari:2018wyi}
\begin{equation}
\begin{aligned}
  \langle V(p_V,\epsilon)|\bar q\gamma_\mu b|\bar B(p_B)\rangle
  &=
  \frac{2V^{B\to V}(q^2)}{m_B+m_V}
  \varepsilon_{\mu\nu\alpha\beta}
  \epsilon^{*\nu}p_B^\alpha p_V^\beta,\\[3pt]
  \langle V(p_V,\epsilon)|\bar q\gamma_\mu\gamma_5 b|
  \bar B(p_B)\rangle
  &=i\Bigg\{
  2m_V A_0^{B\to V}(q^2)
  \frac{\epsilon^*\!\cdot q}{q^2}q_\mu +(m_B+m_V)A_1^{B\to V}(q^2)
  \left[
    \epsilon_\mu^*
    -\frac{\epsilon^*\!\cdot q}{q^2}q_\mu
  \right]\\
  &\quad -A_2^{B\to V}(q^2)
  \frac{\epsilon^*\!\cdot q}{m_B+m_V}
  \left[
    (p_B+p_V)_\mu
    -\frac{m_B^2-m_V^2}{q^2}q_\mu
  \right]
  \Bigg\},
\end{aligned}
\label{eq:BtoV_form_factor_definition}
\end{equation}
where $q=p_B-p_V$, $A_0^{B\to V}(0)=A_3^{B\to V}(0)$, and
\begin{equation}
  A_3^{B\to V}(q^2)
  =
  \frac{m_B+m_V}{2m_V}A_1^{B\to V}(q^2)
  -\frac{m_B-m_V}{2m_V}A_2^{B\to V}(q^2).
\end{equation}

The three-hadron strong couplings entering Eqs.~(B.1)--(B.4) are
defined through the on-shell matrix elements
\begin{equation}
\begin{aligned}
  &\langle \mathcal B_f(p_f)P(p)|i\mathcal L_{\rm str}|
  \mathcal B_i(p_i)\rangle
  =
  g_{P\mathcal B_i\mathcal B_f}\,
  \bar u_f(p_f)i\gamma_5u_i(p_i),\\[3pt]
  &\langle \mathcal B_f(p_f)V(p,\epsilon)|i\mathcal L_{\rm str}|
  \mathcal B_i(p_i)\rangle\\
  &\quad =
  \epsilon^{*\mu}\bar u_f(p_f)
  \left[
    f_{1V\mathcal B_i\mathcal B_f}\gamma_\mu
    -\frac{i f_{2V\mathcal B_i\mathcal B_f}}{m_i+m_f}
    \sigma_{\nu\mu}p^\nu
  \right]
  u_i(p_i),
\end{aligned}
\label{eq:strong_vertex_definitions}
\end{equation}
here $p=p_i-p_f$, while
$g_{P\mathcal B_i\mathcal B_f}$,
$f_{1V\mathcal B_i\mathcal B_f}$, and
$f_{2V\mathcal B_i\mathcal B_f}$
denote the corresponding strong coupling constants.

\clearpage
\bibliographystyle{unsrturl}
\bibliography{References} 

\end{document}